\documentclass[onecolumn]{revtex4}

\usepackage{graphicx}
\usepackage{dcolumn}
\usepackage{bm}

\input epsf

\begin{document}

\title{\Large Emergent Universe in Brane World Scenario}

\author{\bf Asit~Banerjee$^1$, Tanwi~Bandyopadhyay$^2$ and
~Subenoy~Chakraborty$^2$\footnote{schakraborty@math.jdvu.ac.in}}

\affiliation{$^1$Department of Physics, Jadavpur University,
Kolkata-32, India.\\ $^2$Department of Mathematics, Jadavpur
University, Kolkata-32, India.}

\date{\today}

\begin{abstract}
A model of an emergent universe is obtained in brane world. Here
the bulk energy is in the form of cosmological constant, while
the brane consists of the Chaplygin gas with the modified
equation of state such as $p=A\rho-B/\rho$. Initially the brane
matter for the special choice $A=\frac{1}{3}$ may have negative
or positive pressure depending on the relative magnitudes of the
parameter $B$ and the cosmological constant of the bulk, while
asymptotically in future the brane world approaches a
$\Lambda$CDM model.
\end{abstract}

\maketitle

In 1967 Harrison [1] obtained a model of the closed universe
containing radiation, which approaches the state of an Einstein
static model asymptotically, i.e, as $t\rightarrow-\infty$. This
kind of model has so far been discovered subsequently by several
workers in the recent past such as that of Ellis and Maartens
[2], Ellis et al. [3]. They obtained closed universes with a
minimally coupled scalar field $\phi$ with a special form for
self interacting potential and possibly some ordinary matter with
equation of state $p=\omega\rho$ where
$-\frac{1}{3}\leq\omega\leq1$. However, exact analytic solutions
were not presented in these models, although their behaviour alike
that of an emergent universe was highlighted. An emergent
universe is a model universe in which there is no timelike
singularity, is ever existing and having almost static behaviour
in the infinite past ($t\rightarrow-\infty$) as is mentioned
earlier. The model eventually evolves into an inflationary stage.
In fact, the emergent universe scenario can be said to be a
modern version and extension of the original Lemaitre-Eddington
universe. Mukherjee et al. [4] obtained solutions for Starobinsky
model for flat FRW space time and studied the features of an
emergent universe. Very recently, a general framework for an
emergent universe model has been proposed by Mukherjee et al. [5]
using an adhoc equation of state connecting the pressure and
density. However, these solutions require exotic matter in many
cases.\\

In our present paper, we propose an interesting model of an
emergent universe existing in the brane, while the bulk contains
only the 5D cosmological constant. The brane energy content is
that of a modified Chaplygin gas widely discussed in the last few
years [6-9]. The generalized Chaplygin gas has usually the
equation of state [10] $p=-B/\rho^{\alpha}$ ($\alpha=1$
corresponds to Chaplygin gas model [11]), which is modified later
in the form

\begin{equation}
p=A\rho-\frac{B}{\rho^{\alpha}}
\end{equation}

where $A$, $B~(>0)$ and $\alpha~(0<\alpha\leq1)$ are constants.
Such models start with dust or radiation or any other form of a
perfect fluid in the beginning and inflates to a final stage of
$\Lambda$CDM model with the equation of state $p=-\rho$.\\

The geometry of the five dimensional bulk is assumed to be
characterized by the space time metric of the form

\begin{equation}
ds^{2}=-n^{2}(t,y)dt^{2}+a^{2}(t,y)\delta_{ij}dx^{i}dx^{j}+b^{2}(t,y)dy^{2}
\end{equation}

where $y$ is the fifth coordinate and the hypersurface $y=0$ is
identified as the world volume of the brane that forms our
universe. For simplicity, we choose the usual spatial section of
the brane to be flat. Now following Bin\'{e}truy et al. [12,13],
the energy conservation equation on the brane reads

\begin{equation}
\dot{\rho_{b}}+3(\rho_{b}+p_{b})\frac{\dot{a_{0}}}{a_{0}}=0
\end{equation}

which integrating once (using the equation of state (1)) gives
[7,14]

\begin{equation}
\rho_{b}=\left[\frac{1}{(1+A)}\left\{\frac{\rho_{0}}{{a_{0}}^{3(1+A)(1+\alpha)}}
+B\right\}\right]^{1/(1+\alpha)}
\end{equation}

($\rho_{0}$, an arbitrary integration constant)\\

Using this form for $\rho_{b}$, the generalized Friedmann type
equations take the form (see eqns. (45) and (46) in ref. [14])

\begin{equation}
\frac{\dot{a_{0}^{2}}}{a_{0}^{2}}=\frac{\kappa^{2}\rho_{b}^{2}}{36}
+\frac{\kappa^{2}\Lambda_{5}}{6}+\frac{C}{a_{0}^{4}}
\end{equation}

and

\begin{equation}
\ddot{a_{0}}=-\frac{\kappa^{2}}{36}\frac{(3A+2)}{(A+1)}\frac{\rho_{0}}{a_{0}^{(6A+5)}}
+\frac{\kappa^{4}B}{36~(1+A)}~a_{0}+\frac{\kappa^{2}\Lambda_{5}}{6}~a_{0}-\frac{C}{a_{0}^{3}}
\end{equation}

with $C$ being an integration constant.\\

We note that for positive cosmological constant
($\Lambda_{5}>0$), there is a transition from deceleration to
acceleration with only a maximum but no minimum. It is a
recollapsing brane model. But for negative cosmological constant,
either there is deceleration throughout or a transition from
deceleration to acceleration depending on the magnitude of
$\Lambda_{5}$ (i.e, $|\Lambda_{5}|>$ or
$<\frac{\kappa^{2}B}{6(1+A)}$) (for detail discussion see ref. [14]).\\

We now proceed to solve equation (6) for $A=\frac{1}{3}$ (for
which solution in closed form is possible). The first integral of
equation (6) can be written in an integral form as

\begin{equation}
\frac{1}{4}\int\frac{du}{\sqrt{bu^{2}+Cu+d}}=\pm~(t-t_{0})
\end{equation}

with $u=a_{0}^{4},~b=\frac{\kappa^{4}B}{48}
+\frac{\kappa^{2}\Lambda_{5}}{6},~d=\frac{\kappa^{4}\rho_{0}}{48}$.\\

The explicit solution is given by ($b>0$)

\begin{equation}
a_{0}^{4}=\left\{
\begin{array}{ll}
\frac{\sqrt{4bd-C^{2}}}{2b}~Sinh\left[\pm4\sqrt{b}~(t-t_{0})\right]-\frac{C}{2b},
~~~~~~~~~~~~~~~~(\text{when $4bd>C^{2}$})\\\\
\frac{\sqrt{C^{2}-4bd}}{2b}~Cosh\left[4\sqrt{b}~(t-t_{0})\right]-\frac{C}{2b},
~~~~~~~~~~~~~~~~~~(\text{when $4bd<C^{2}$})\\\\
\frac{1}{2b}\left[e^{\pm4\sqrt{b}~(t-t_{0})}\right]-\frac{C}{2b},
~~~~~~~~~~~~~~~~~~~~~~~~~~~~~~~~~~~(\text{when $4bd=C^{2}$})
\end{array}
\right.
\end{equation}

We note that for $C>0$, all the above solutions start from big
bang singularity and expands indefinitely as
$t\rightarrow\infty$. However, for $C<0$, the behaviour of the
first solution remains same while the second solution represents
a bouncing solution having minimum at finite time. The third
solution is a singularity free solution, starting with finite
$a_{0}$ at $t=-\infty$ (where both $\dot{a_{0}}$ and
$\ddot{a_{0}}$ vanish) and expands exponentially. The brane model
corresponding to this solution is termed as emergent universe
[1-3] in brane world scenario.\\

Further, for $b<0$ (a negative cosmological constant with
$|\Lambda_{5}|>\frac{\kappa^{2}B}{8}$), the solution can be
written as

\begin{equation}
a_{0}^{4}=\frac{C}{2|b|}+\frac{\sqrt{4|b|d+C^{2}}}{2|b|}\stackrel{Sin}{
\stackrel{or}{Cos}}[4\sqrt{|b|}~(t-t_{0})]
\end{equation}

Also for $b=0$, i.e, negative cosmological constant having
magnitude fine tuned to $\frac{\kappa^{2}B}{8}$, the solution
takes the form

\begin{equation}
a_{0}^{4}=\left\{
\begin{array}{ll}
4C(t-t_{0})^{2}-\frac{d}{C}\\\\
~~~~~~~~\text{or}\\\\
\frac{d}{|C|}-4|C|(t-t_{0})^{2}
\end{array}
\right.
\end{equation}

according as $C>0$ or $<0$.\\

These solutions have familiar behaviour and are not of much
interest in the present context.\\

We shall now discuss the properties of that solution in equations
(8) representing a model for emergent universe. Asymptotically in
the past (i.e, at $t\rightarrow-\infty$) the scale factor $a_{0}$
has the constant value $(\frac{|C|}{2b})^{1/4}$ and using equation
(4), the matter density has initially the constant value

\begin{equation}
\rho_{bi}=\left(\frac{3B}{2}+\frac{6\Lambda_{5}}{\kappa^{2}}\right)^{1/2}
\end{equation}

Consequently, the equation of state, i.e, equation (1), in the
asymptotic past, takes the form

\begin{equation}
p_{bi}=-\frac{1}{3}~\rho_{bi}+\frac{4\Lambda_{5}}{\kappa^{2}\rho_{bi}}
\end{equation}

If the bulk cosmological constant $\Lambda_{5}$ is negative (with
$|\Lambda_{5}|<\frac{\kappa^{2}B}{4}$), then pressure in the
brane matter is negative throughout the evolution and finally
goes to $\Lambda$CDM model. On the other hand, for positive
cosmological constant in the bulk, i.e, for $\Lambda_{5}>0$, the
initial pressure $p$ may be positive or negative depending on the
relative magnitude of $\Lambda_{5}$ and $B$. In order that the
initial pressure $p_{bi}$ is positive for $\Lambda_{5}>0$, we
must have $\rho_{bi}^{2}<\frac{12\Lambda_{5}}{\kappa^{2}}$, which
in turn demands $B<\frac{4\Lambda_{5}}{\kappa^{2}}$. The Chaplygin
gas brane therefore starts in this case with initial positive
pressure and eventually evolves to the $\Lambda$CDM model, which
is the characteristic of all Chaplygin gas
cosmological models.\\

The other point to be noted in the present emergent brane model
is that, here the spatial curvature is zero unlike the closed
universes considered in many of the previous emergent models.\\

{\bf Acknowledgement}:\\

TB is thankful to CSIR, Govt. of India, for
awarding JRF.\\

\end{document}